\documentclass[lettersize,journal]{IEEEtran}
\usepackage{amsmath,amsfonts}

\usepackage{algorithmic}
\usepackage{algorithm}
\usepackage{array}
\usepackage[colorlinks=true,linkcolor=red,citecolor=green,urlcolor=blue]{hyperref}
\usepackage{textcomp}
\usepackage{stfloats}
\usepackage{url}
\usepackage{verbatim}
\usepackage{graphicx}
\usepackage{cite}
\usepackage{soul}
\usepackage{makecell}
\usepackage{subcaption}
\usepackage{amssymb}

\begin{document}

\title{Start from Video-Music Retrieval: An Inter-Intra Modal Loss for Cross Modal Retrieval}

\author{Zeyu Chen, Pengfei Zhang, Kai Ye, Wei Dong, Xin Feng, Yana Zhang
}

\maketitle

\begin{abstract}
The burgeoning short video industry has accelerated the advancement of video-music retrieval technology, assisting content creators in selecting appropriate music for their videos. In self-supervised training for video-to-music retrieval, the video and music samples in the dataset are separated from the same video work, so they are all one-to-one matches. This does not match the real situation. In reality, a video can use different music as background music, and a music can be used as background music for different videos. Many videos and music that are not in a pair may be compatible, leading to false negative noise in the dataset. A novel inter-intra modal (II) loss is proposed as a solution. By reducing the variation of feature distribution within the two modalities before and after the encoder, II loss can reduce the model’s overfitting to such noise without removing it in a costly and laborious way. The video-music retrieval framework, II-CLVM (Contrastive Learning for Video-Music Retrieval), incorporating the II Loss, achieves state-of-the-art performance on the YouTube8M dataset. The framework II-CLVTM shows better performance when retrieving music using multi-modal video information (such as text in videos). Experiments are designed to show that II loss can effectively alleviate the problem of false negative noise in retrieval tasks. Experiments also show that II loss improves various self-supervised and supervised uni-modal and cross-modal retrieval tasks, and can obtain good retrieval models with a small amount of training samples. 
\end{abstract}


\section{Introduction}

With the increasing content demands in the short video industry, AI-assisted video editing has greatly increased the efficiency in video production. To choose a piece of good background music (BGM) for a video by AI is our main research point. Some supervised learning based algorithms select music by the matching scores of tags \cite{4:4} \cite{25:25} or the feature distance in the emotional space \cite{1:1} \cite{5:5} \cite{6:6}. In recent works \cite{18:18} \cite{19:19} \cite{22:22}, music selection becomes a cross-modal retrieval task based on contrastive learning. Video and audio encoders learn various of video-music matching factors in a self-supervised training way. The cosine distance of the encoded cross-modal features is used to assess the suitability of media candidates.

This paper focuses on solving the problem of training with noisy data in self-supervised cross-modal retrieval. In self-supervised learning, a pair of video-music samples come from the same video work, so the dataset has only one-to-one matches. This obviously does not accurately reflect the real situation. In fact, a piece of music can be used as the background music for different videos, and a video can also be paired with different background music. In this case, there are many suitable videos and music that are not in the same pair, resulting in many false negative noisy samples. The cross-modal training objective is to minimize the distance between positive samples and maximize the distance between negative samples. When the model is overfitted to the noise, the distance between many false negative sample pairs is maximized, leading to a decrease in the model’s generalization ability. To tackle this challenge, a novel inter-intra modal loss (II Loss) is specifically designed to handle this type of noise. The II Loss addresses the issue by using the intra loss component to minimize drastic variations in the feature distributions within each modality during training. This approach effectively mitigates overfitting on the noisy data and allows for more accurate retrieval of relevant matches without requiring complex noise removal techniques. The proposed framework Inter-Intra Contrastive Learning for Video-Music Retrieval (II-CLVM) based on inter-intra modal loss achieves the state-of-the-art on video-music retrieval on Youtube8M and performs better when retrieving music using multi-modal video information (such as text). II loss is also performs well for other cross-modal retrieval tasks.

Our contributions are as follows:

\begin{itemize}
\item This paper proposes inter-intra modal loss (II loss), which enables the retrieval models trained on noisy data to have better generalization ability. II loss alleviates the model’s overfitting to false negative noise by minimizing the drastic changes in feature distribution within each modality. II loss works well in various self-supervised and supervised uni-modal and cross-modal retrieval tasks.
\item The II-CLVM video-music retrieval framework is developed. It incorporates II Loss and achieves state-of-the-art performance on the YouTube8M dataset. Additionally, the framework employs Global Sparse (GS) sampling which allows music retrieval to be based on the content of the complete video, rather than on fixed-duration video clips. The framework can also easily integrate multi-modal video information (such as images and text) to achieve better performance.
\end{itemize}

\begin{figure*}[ht]
\centering
\includegraphics[scale=0.53]{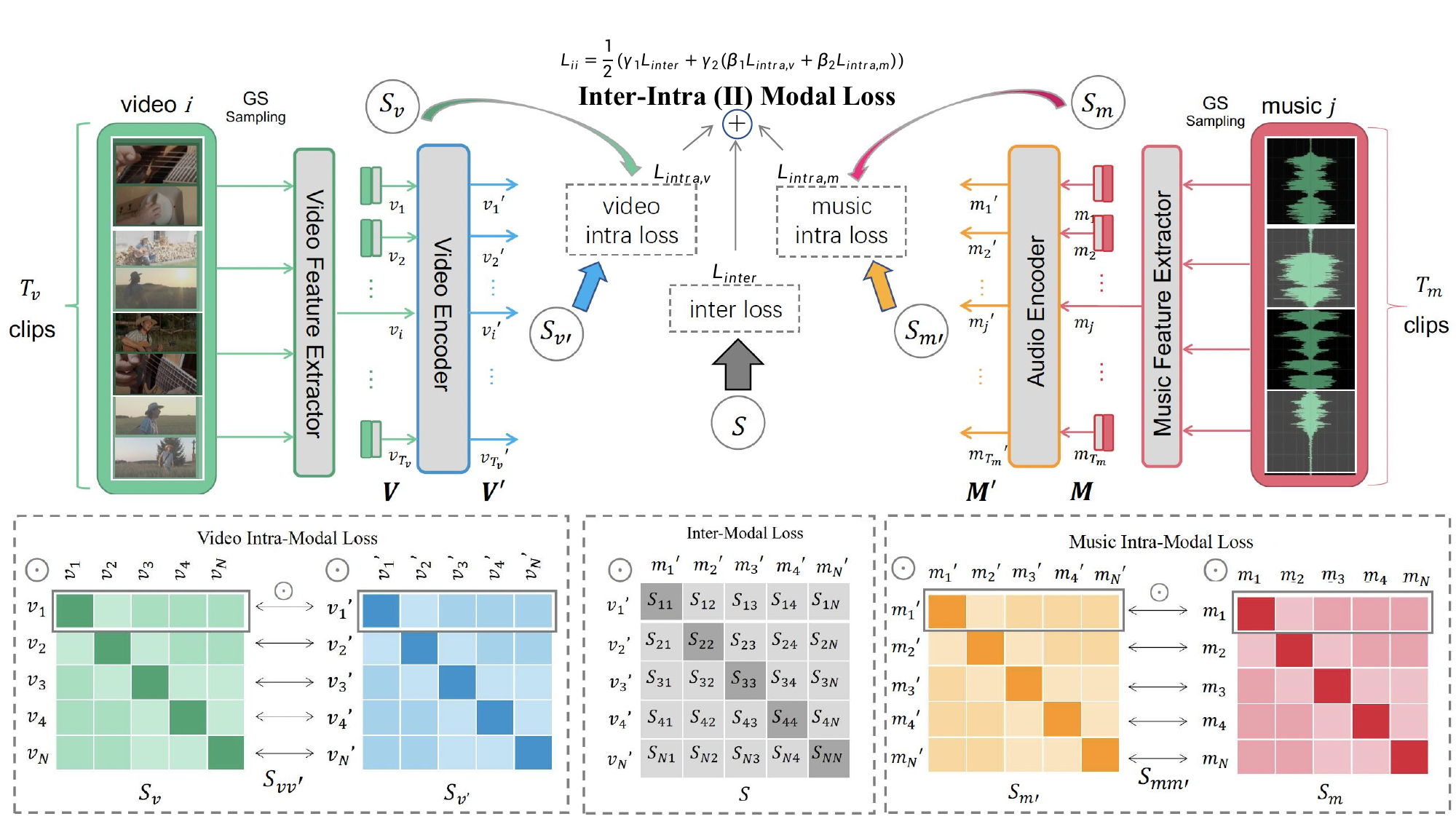} %
\caption{The structure of II-CLVM. The global sparse (GS) sampling method is applied for each video and music to extract the pretrained feature sequences $v_{i}$ and $a_{j}$. The encoded features $v_{i}'$ and $m_{j}'$ are then obtained by video and music encoders, respectively. Then, the inter-modal similarity matrix $S$ and the intra-modal similarity matrix $S_{v}$, $S_{v'}$, $S_{m}$, $S_{m'}$ are calculated. The inter-modal loss is calculated from the matrix $S$, and the intra-modal losses for video and music modalities are calculated by $S_{v}$ and $S_{v'}$, $S_{m}$ and $S_{m'}$, respectively.}
\label{fig:II-CLVM}
\end{figure*}

\section{Related Work}

In recent works, background music selection for video is a cross-modal retrieval task based on pretrained features. Contrastive learning enables the model to learn rich matching rules in large-scale video-music datasets and reduce the feature distance between matched video-music pairs.

\subsection{Contrastive learning and cross-modal retrieval}

Due to the pervasively existed video music pairs online, retrieving a music for a video is usually treated as a cross-modal contrastive learning task. Early works such as \cite{22:22} performed binary classification for pairwise cross-modal samples. Ranking/triplet loss ordered the similarity score of samples when the query was given (e.g. \cite{19:19},\cite{22:22},\cite{36:36}). The distance or similarity of hash codes \cite{15:15} \cite{16:16}\cite{55:55hash}, the canonical correlation analysis (CCA) \cite{23:23} and the relational network (RN) \cite{14:14} were also applied to model the cross-modal distance. In recent works, CLIP (Contrastive Language–Image Pre-training)\cite{24:24} used a large number of image-text pairs to self-supervise the pre-training of two encoders for image-text retrieval tasks. During pre-training, the distance between cross-modal feature pairs of each mini-batch was processed by a simple cross entropy loss function. CLIP was then extended to various types of cross-modal retrieval and pre-training tasks such as video-text (CLIP4clip\cite{31:31}) and audio-text (AudioCLIP\cite{37:37}). In addition to matching and retrieval tasks, some recent works (e.g. BLIP\cite{48:48blip}) explored different pre-training methods that enable encoders to migrate well to caption and QA tasks.

There are also many works that try to optimize the cross modal retrieval task by redesigning of loss function. Some works \cite{18:18} \cite{21:21} integrated the label classification loss into the inter-modal similarity loss. In order to prevent intra-modal structure collapse during cross-modal training, \cite{19:19} and \cite{59:59itc} added intra-modal loss to triplet loss. \cite{12:12} designed a three-part loss for robust cross-modal retrieval.

\subsection{BGM Selection for videos}

To find an appropriate BGM from a music library based on the contents of both music and the given video, early works were more likely to choose music by matching labels. Emotion was always regarded as a critical factor in music-video retrieval \cite{1:1} \cite{6:6}. In addition, some works also incorporated factors such as location (\cite{4:4} \cite{5:5}), style and genre (\cite{8:8}), comments and lyrics text (\cite{17:17}), and user personality (\cite{5:5}). In recent works, algorithms \cite{18:18} \cite{19:19} \cite{21:21} \cite{22:22} are trained by self-supervised contrastive learning methods and capable to learn the rich factors and preferences of video creators to choose BGM automatically. There are also some works (\cite{50:50generate1} \cite{51:51generate2}) generated new BGM for videos.

\subsection{Pretrained visual and audio feature extractors}

In video-music retrieval task, most works retrieved music based on the distance between the pretrained visual and audio features. The performance of pretrained feature extractors also becomes a key factor restricting retrieval performance. These pretrained feature extractors are pre-trained on large-scale datasets, and all parameters are frozen during fine-tuning. When extracting video embeddings, multiple frames can be sampled and the frame-level features such as ResNet-50 \cite{27:27}, Inception-V3 \cite{28:28}, ViT \cite{32:32}, or CLIP-Vision\cite{24:24} can be extracted. Also, video-level features like I3D\cite{52:52i3d}, SlowFast\cite{53:53slowfast}, and Video Swin Transformer\cite{47:47vidswin} can be extracted. As for audio, frame-level features VGGish \cite{29:29} or PANNs \cite{30:30} can be extracted.

\section{The proposed framework}

Fig \ref{fig:II-CLVM} shows the architecture of the framework II-CLVM (Inter-Intra Contrastive Learning for Video-Music Retrieval) with the proposed inter-intra modal loss. During the model training, there are $N$ video-music pairs in each mini-batch. Firstly, global sparse (GS) sampling is performed on both video and music, and extract the pretrained feature sequences $\textbf{V}=\left \{ v_{i}  \right \} _{i=1}^{N}$ for video and $\textbf{M}=\left \{ m_{j}  \right \} _{j=1}^{N}$ for music. Then, the video embeddings $\textbf{V'}=\left \{ v_{i}'  \right \} _{i=1}^{N}$ and music embeddings $\textbf{M'}=\left \{ m_{j}'  \right \} _{j=1}^{N}$ are obtained by separate sequence encoders. The inter-intra (II) modal loss is proposed to measure the distance between the encoded video embeddings and music embeddings. The detail of each module of II-CLVM is introduced below.

\subsection{Global sparse sampling}

Existing video-music retrieval usually takes one continuous fixed-duration (FD) clip from the original media to represent the whole sequence, e.g. cutting 30$s$ around the center of both video and music as in \cite{19:19}. Those methods ignore the rest parts of video and music, so that the retrieved music may only be partially related to the video. To extract features of the entire video and the whole music, the global sparse (GS) sampling \cite{33:33} is applied. For video $i$, it is split evenly into $T_{v}$ clips and the video feature sequence $v_{i} \in \mathbb{R}^{T_{v}\times E_{v}}$ is obtained where $E_{v}$ is the dimension of the feature. Similarly, the audio feature sequence $m_{j} \in \mathbb{R}^{T_{m}\times E_{m}}$ is obtained for music $j$. Note that the purpose of extracting feature sequences of fixed length for video and music of different durations is to eliminate duration information and enable the model to retrieve based on content.

\subsection{Sequence encoders}

To extract the temporal information from the frame-level video and music feature sequences, $\textbf{V}$ and $\textbf{M}$ are fed into two sequence encoders (biLSTM, transformer encoder, etc), respectively. After encoding, the encoded video feature $\textbf{V'}=\left \{ v_{i}'  \right \} _{i=1}^{N},\;({v_{i}}' \in \mathbb{R}^{1 \times D})$ and music feature $\textbf{M'}=\left \{ m_{j}'  \right \} _{j=1}^{N},\;({m_{j}}' \in \mathbb{R}^{1 \times D})$ are obtained, where $D$ is the fixed hidden dimension of the sequence encoders for both video and music modalities.

\subsection{The inter-intra (II) modal loss}

\begin{figure}[t]
\centering
\includegraphics[scale=0.25]{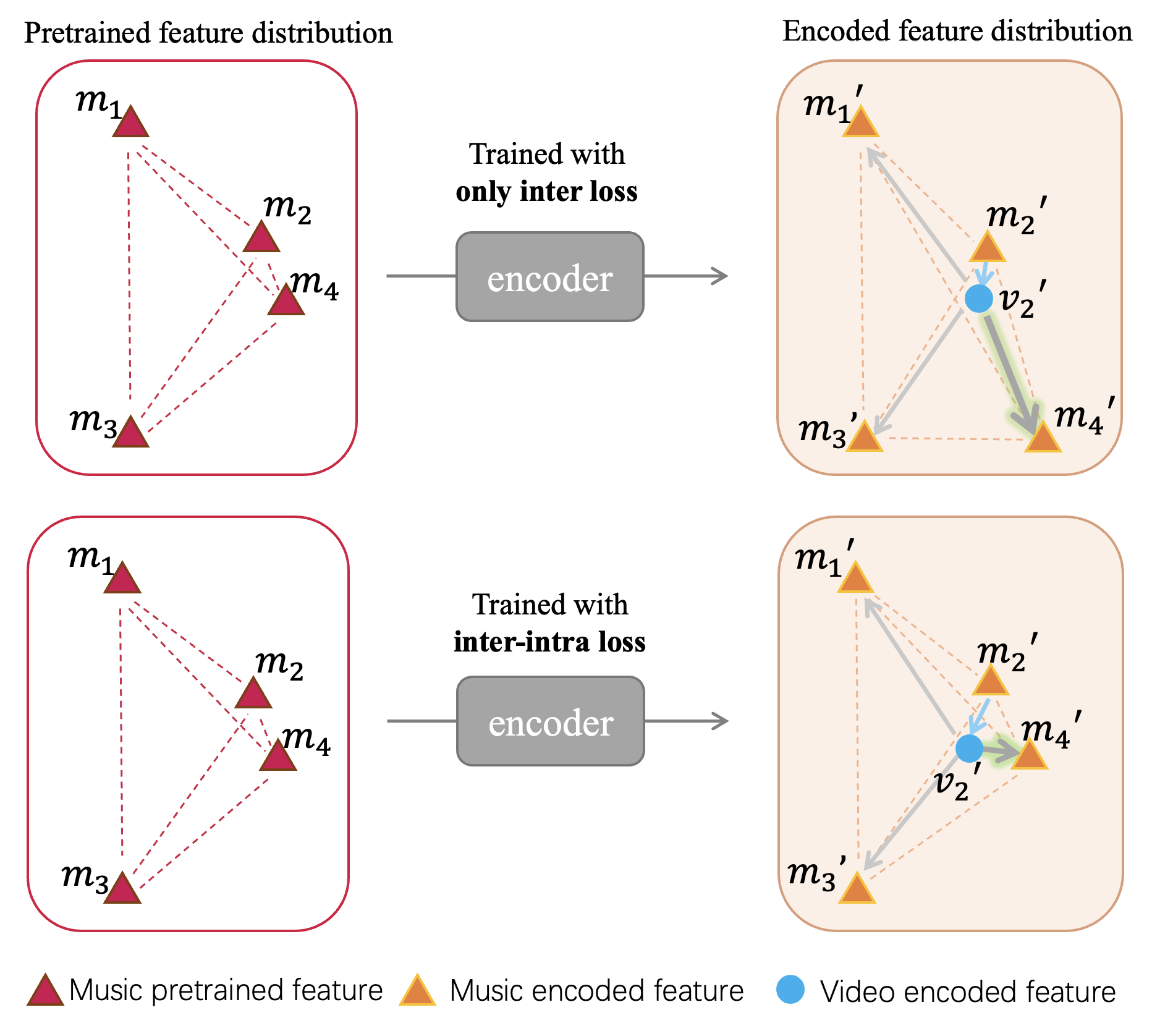} %
\caption{Music feature distribution before and after encoder. The dotted line indicates the feature distribution within a batch. The solid arrows represent the direction in which inter-modal loss acts on the features. The gray arrow increases the feature distance and the blue arrow decreases the distance. Intra-modal loss prevents the distance of false negative sample pairs ($v_2$ and $m_4$) from being larger by maintaining the distribution of pretrained features.}
\label{fig:point}
\end{figure}

As illustrated in Fig \ref{fig:point}, we consider an example of a batch with a size of 4. The video $v_2$ and the music $m_2$ constitute a pair of positive samples. Due to the similarity between music $m_4$ and $m_2$, $m_4$ can also be effectively utilized as background music (BGM) for video $v_2$. During cross-modal training, if solely relying on the conventional inter-modal loss, the distance between the features output by the encoder $v_{2}'$ and $m_{2}'$ would decrease, while the distance between the features $v_{2}'$ and $m_{4}'$ would increase, which is an undesired outcome. As depicted in Fig \ref{fig:point}, the proposed intra loss aims to preserve the relative feature distribution of the four music samples in the encoder output, as close as possible to their pre-encoder state, thereby preventing $m_{4}'$ from moving away from $v_{2}'$.

In II-CLVM, a mini-batch consists of $N$ matched video-music pairs. In each batch, the loss is a weighted sum of the inter-modal similarity loss and the intra-modal features distance distribution loss as shown in Fig \ref{fig:II-CLVM}.

\subsubsection{\textbf{Inter-modal loss}}

As shown in equation (\ref{eq:inter}), the inter-modal loss is calculated based on the cosine similarity matrix $S \in \mathbb{R}^{N\times N}$. Each element of $S$ is calculated as follows:

\begin{equation}
S_{(i,j)}={v_{i}}' \circledast {m_{j}'},\;(i,j\in \left\{1,2,...,N \right\})
\end{equation}

where $\circledast$ is defined as the cosine similarity between two vectors.

\begin{equation}
\begin{aligned}
L_{inter}=\frac{1}{N}(\alpha_{1}\sum_{i=1}^{N}\textbf{CE}(\sigma(e^{n_{t}}S_{i,:}),I_{i,:})\\
+\alpha_{2}\sum_{j=1}^{N}\textbf{CE}(\sigma(e^{n_{t}}S_{:,j}),I_{:,j}))
\end{aligned}
\label{eq:inter}
\end{equation}

In $S$, $S_{i,:}$ is the $i$th row, and $S_{:,j}$ is the $j$th column. $n_{t}$ is a learnable temperature parameter and $e^{n_{t}}$ controls the range of the logits in the Softmax function. $I$ is an $N$-order identity matrix. $\textbf{CE}(\cdot )$ is the cross entropy loss, and $\sigma(\cdot )$ means the Softmax function. $\alpha_{1}$ and $\alpha_{2}$ are adjustable parameters. The inter-modal loss increases values on the diagonal of $S$ and decreases those in other positions.

\subsubsection{\textbf{Intra-modal loss}}

For the video modality, two intra-modal similarity matrices $S_{v} \in \mathbb{R}^{N\times N}$ and $S_{v'} \in \mathbb{R}^{N\times N}$ are calculated as shown in Fig \ref{fig:II-CLVM}. In a mini-batch, $S_{v}$ and $S_{{v}'}$ describe the similarity of different video features before and after the encoder, respectively. Here, each element of $S_{v}$ and $S_{{v}'}$ is calculated as follows:

\begin{equation}
S_{v(i,j)}=\bar{v}_i \circledast \bar{v}_j,\;(i,j\in \left\{1,2,...,N \right\})
\label{eq:sv}
\end{equation}

\begin{equation}
S_{v'(i,j)}=v_{i}' \circledast v_{j}',\;(i,j\in \left\{1,2,...,N \right\})
\label{eq:sv'}
\end{equation}

where $\bar{v}_i$ is the temporal average of $v_i$. To achieve the invariance of feature distribution before and after encoding, $S_{v}$ and $S_{{v}'}$ should be similar. The vectors $S_{vv'\_r} \in \mathbb{R}^{1 \times N}$ and $S_{vv'\_c} \in \mathbb{R}^{1 \times N}$ are calculated to describe the row similarity and column similarity between $S_{v}$ and $S_{{v}'}$, respectively. The calculation is as follows:

\begin{equation}
S_{vv'\_r,i}=S_{v,i,:} \circledast S_{v',i,:},\;(i\in \left\{1,2,...,N \right\})
\end{equation}

\begin{equation}
S_{vv'\_c,j}=S_{v,:,j} \circledast S_{v',:,j},\;(j\in \left\{1,2,...,N \right\})
\end{equation}

Then, the intra-modal loss is calculated as follows:

\begin{equation}
\begin{aligned}
L_{intra,v}=\frac{\delta_{1}}{N}\sum_{i=1}^{N}(1-S_{vv'\_r,i})\\
+\frac{\delta_{2}}{N}\sum_{j=1}^{N}(1-S_{vv'\_c,j})
\end{aligned}
\label{eq:intrav1}
\end{equation}

where $\delta_{1}$ and $\delta_{2}$ are the weight parameters. Here, $S_{v(i,j)}=S_{v(j,i)}$, $S_{v'(i,j)}=S_{v'(j,i)}$, so $S_{vv'\_r}$ is equivalent to $S_{vv'\_c}$, denoted as $S_{vv'}$. The intra-modal loss can be simplified to:

\begin{equation}
L_{intra,v}=\frac{1}{N}\sum_{i=1}^{N}(1-S_{vv',i})
\label{eq:intrav2}
\end{equation}

The music modality is processed in the same way. The total intra-modal loss $L_{intra}$ is computed as equation (\ref{eq:intra}), where $\beta_{1}$ and $\beta_{2}$ are the weights of intra loss of the two modalities respectively. The inter-intra modal loss $L_{ii}$ is a weighted sum of inter-modal loss and intra-modal loss as shown in equation (\ref{eq:ii}). $L_{ii}$ could work with various feature extractors and both of the above sampling methods.

\begin{equation}
L_{intra}=\beta_{1}L_{intra,v}+\beta_{2}L_{intra,m}
\label{eq:intra}
\end{equation}

\begin{equation}
L_{ii}=\frac{1}{2}(\gamma_{1}L_{inter}+\gamma_{2}L_{intra})
\label{eq:ii}
\end{equation}

\begin{figure}
\centering
\includegraphics[width=0.99\columnwidth]{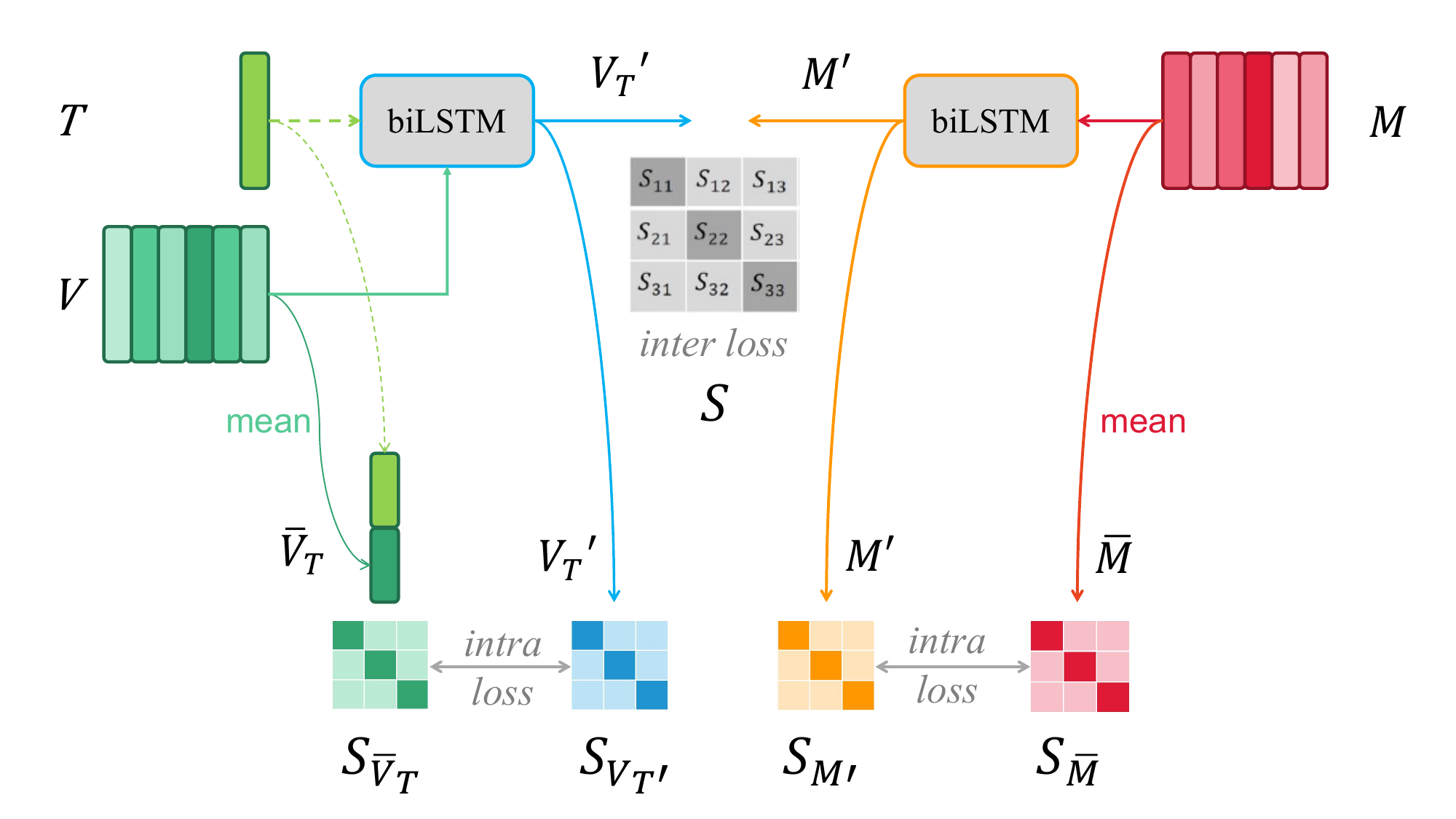} %
\caption{The general structure of II-CLVTM framework. Feature $V_{T}'$ is the result of encoder fusion of the video feature sequence $V$ and the text feature sequence $T$. Feature $M$ is processed by another encoder to obtain feature $M'$. The inter-modal similarity matrix $S$ is calculated from $V_{T}'$ and $M'$. The intra-modal similarity matrices $S_{\bar{V}_T}$), $S_{V_{T}'}$, $S_{\bar{M}}$ and $S_{M}'$ are obtained from $\bar{V}_T$(concatenated from $T$ and $\bar{V}$), $V_T'$, $\bar{M}$ and $M'$ respectively. The inter loss is calculated with $S$. The intra loss is calculated with $S_{\bar{V}_T}$), $S_{V_{T}'}$, $S_{\bar{M}}$ and $S_{M}'$, respectively.}
\label{fig:iiclvtm}
\end{figure}

\subsection{II-CLVTM}

When choosing a background music (BGM) for a video, the video creator may also provide some text information, such as the video’s description, title or keywords. Our framework can easily incorporate these text inputs. We use the Inter-Intra Contrastive Learning for Video$\&$Text-to-Music Retrieval (II-CLVTM) model, as shown in Fig \ref{fig:iiclvtm}, to fuse and encode the raw video feature sequences $V$ and text feature vectors $T$ with biLSTM. The text feature vector is used as the initial hidden vector for biLSTM. As Fig \ref{fig:iiclvtm} shows, when the query is multi-modal, we compute the intra-modal loss based on the similarity matrix of the features before and after the cross-modal encoder. We also note that the uncoded feature $\bar{V}_T$ is obtained by concatenating the raw text feature and the average video feature.

\section{Experiments and results}

The experiments in this section consist of three parts. Section \ref{sec:youtubeexp} mainly introduces the experiments of II-CLVM on the video-music retrieval task on the YouTube8M dataset. Section \ref{sec:otherexp} verifies the generality of II loss on various other cross-modal retrieval tasks. Section \ref{sec:thirdexp} designs two experiments to verify that II loss can effectively alleviate the problem of false negative noise in the retrieval datasets.

\subsection{Experiments on YouTube8M}\label{sec:youtubeexp}

In this section, the performance of each module in II-CLVM is tested on video-music retrieval task of official YouTube8M dataset\cite{9:9}. What's more, a subjective evaluation is conducted to check the BGM selection performance of II-CLVM on YouTube8M.

In YouTube8M \cite{9:9}, there are 149,213 video samples labeled as Music Video, in which 116,098 videos are for training and 33,115 videos are for testing. Three testing sets $A_{1024}, B_{1000}$ with 1024, and 1000 pairs are randomly generated from the test data. YouTube8M provides the official video features (Inception-v3 \cite{28:28}, 1024-d per second after whitening PCA) and music features (VGGish \cite{29:29}, 128-d per second). In the following experiments, these features are used as the raw pretrained features.

All experiments are conducted in a 32G Tesla V100 GPU accelerated environment. The model is built with PyTorch. In II-CLVM, $N=32$, $\alpha_{1}=\alpha_{2}=\beta_{1}=\beta_{2}=0.5$, and $\gamma_{1}=1,\gamma_{2}=3$. The initial value of $n_{t}$ is 0.07. When the encoder is biLSTM encoder, the initial value of $n_{t1}$ is also 0.07. When GS sampling method is applied, $T_{v}=T_{m}=100$. Since YouTube8M provides second-level pretrained feature sequences, GS sampling is directly applied on the feature. Each model is trained for 30 epochs on the training dataset and the $recall@k$ ($R@k$) metric is measured on three testing sets $A_{1024}, B_{1000}$, respectively. $R@k$ is the most common evaluation metric in all kinds of retrieval tasks and is the abbreviation for recall at $k$-th in the ranking list, defined as the proportion of correct matchings in top-$k$ retrieved results. Specifically, in video-to-music retrieval, $R@k$ refers to the probability that the correct music is included in the top $k$ retrieved music results. Apart from $R@1$, $R@10$ and $R@25$ are also noteworthy due to the fuzziness of video-music retrieval rule.

\begin{table}[t]
\caption{Results on video-to-music retrieval task using different sampling methods.}
\centering
\begin{tabular}{|l|lll|}
\hline
Group  & $R@1$ & $R@10$ & $R@25$ \\
\hline
biLSTM(CLVM)(FD) & 5.8$\%$  & 31.0$\%$  & 46.7$\%$  \\
biLSTM(II-CLVM)(FD) & \bf{7.5$\%$}   & \bf{31.6$\%$}  & \bf{49.8$\%$}  \\
\hline
biLSTM(CLVM)(GS) & 18.4$\%$   & 54.0$\%$  & 70.2$\%$  \\
biLSTM(II-CLVM)(GS) & \hl{\bf{22.1$\%$}}   & \hl{\bf{55.1$\%$}} & \hl{\bf{70.4$\%$}}  \\
\hline
\end{tabular}
\label{tab:gs}
\end{table}

\begin{table}[t]\footnotesize
\caption{Results of the video-music retrieval. For the video-music retrieval task, the sequential models such as self-attention(Att) and biLSTM perform better than the FC layer and GCN. The inter-intra modal loss works for each type of encoder. Compared with other works, our II-CLVM achieves state-of-the-art results.}
\centering
\begin{subtable}[t]{1.0\linewidth}
\centering
\resizebox{1.0\columnwidth}{!}{
\begin{tabular}{|l|lll|lll|}
\hline
Group  & $R@1$ & $R@10$ & $R@25$ &$R@1$ & $R@10$ & $R@25$ \\
\hline
& \multicolumn{3}{c}{from 1024} & \multicolumn{3}{c}{from 1000} \\
\hline
CME\cite{18:18} & 10.2$\%$  & 40.3$\%$  & - &  - &  - & - \\
CBVMR\cite{19:19} &  -   &  -   & - &8.2$\%$ & 23.3$\%$ & 35.7$\%$ \\
\hline

FC(CLVM)& 7.7$\%$&\bf{31.7$\%$}&48.4$\%$&6.9$\%$&31.5$\%$&49.3$\%$\\
FC(II-CLVM)& \bf{7.8$\%$}&31.6$\%$&\bf{49.2$\%$} &\bf{8.2$\%$} &\bf{32.8$\%$} &\bf{49.9$\%$} \\
\hline
GCN(CLVM)& 5.4$\%$ & 24.8$\%$ & 39.5$\%$ & 3.2$\%$ & 21.9$\%$ & \bf{41.3$\%$}  \\
GCN(II-CLVM)& \bf{5.6$\%$} & \bf{26.1$\%$} & \bf{42.7$\%$} & \bf{4.8$\%$} & \bf{24.4$\%$} & 40.9$\%$  \\
\hline
Att(CLVM)& 4.0$\%$ & 26.4$\%$ & 43.2$\%$ & 5.5$\%$ & 27.3$\%$ & 46.6$\%$ \\
Att(II-CLVM)& \bf{10.4$\%$} & \bf{28.6$\%$} & \bf{56.0$\%$} & \bf{11.4$\%$} & \bf{42.5$\%$} & \bf{58.8$\%$}  \\
\hline
biLSTM(CLVM)& 18.4$\%$ & 52.3$\%$ & 69.2$\%$ & 18.4$\%$ & 54.0$\%$ & 70.2$\%$ \\
biLSTM(II-CLVM)& \hl{\bf{20.5$\%$}} & \hl{\bf{56.1$\%$}} & \hl{\bf{69.6$\%$}} & \hl{\bf{22.1$\%$}} & \hl{\bf{55.1$\%$}} & \hl{\bf{70.4$\%$}}\\
\hline
\end{tabular}}
\caption{Video-to-music retrieval.}
\end{subtable}

\qquad

\begin{subtable}[t]{1.0\linewidth}
\centering
\resizebox{1.0\columnwidth}{!}{
\begin{tabular}{|l|lll|lll|}
\hline
Group  & $R@1$ & $R@10$ & $R@25$ &$R@1$ & $R@10$ & $R@25$ \\
\hline
& \multicolumn{3}{c}{from 1024} & \multicolumn{3}{c}{from 1000}  \\
\hline
CME\cite{18:18} & 9.8$\%$  & 39.6$\%$  & - &  - &  - & - \\
CBVMR\cite{19:19} &  -   &  -   & - &8.9$\%$ & 25.2$\%$ & 37.9$\%$\\
\hline
biLSTM(CLVM)& 18.5$\%$ & 52.2$\%$ & 67.8$\%$ & 19.9$\%$ & 53.9$\%$ & 67.3$\%$ \\
biLSTM(II-CLVM)& \hl{\bf{21.3$\%$}} & \hl{\bf{53.5$\%$}} & \hl{\bf{69.7$\%$}} & \hl{\bf{20.7$\%$}} & \hl{\bf{54.9$\%$}} & \hl{\bf{69.9$\%$}}  \\
\hline
\end{tabular}}
\caption{Music-to-video retrieval}
\end{subtable}

\label{tab:videomusic}
\end{table}

\subsubsection{\textbf{GS sampling}}

Taking biLSTM as an example of the encoder, the FD (fixed-duration) and GS (global sparse) sampling methods are tested on video-to-music retrieval task on testing set $B_{1000}$. The results of Table \ref{tab:gs} show that GS sampling significantly improves the retrieval performance, and the $R@1$ increases from 7.5$\%$ to 22.1$\%$. The FD method can only retrieve music by fixed-duration video clips, and the performance is not as good as the GS sampling method that retrieves by the complete video. 

\subsubsection{\textbf{The sequence encoder}}

In this subsection, different types of encoders are compared on video-to-music retrieval task. The GS sampling method is applied in all experimental groups. The results in Table \ref{tab:videomusic} show that FC (Fully Connected) layer and GCN have relatively poor performance on the video-music retrieval task which involves two modalities that are both time series, because they do not consider temporal information. On the other hand, sequence models such as self-attention(Att) and biLSTM have much better retrieval performance. For example, in the FROM 1024 retrieval, the two sequence models increase the $R@1$ of video-to-music retrieval to 10.4$\%$ and 20.5$\%$, respectively.

\subsubsection{\textbf{II loss in II-CLVM}}

In this subsection, the impact of the II loss is evaluated. As shown in Table \ref{tab:gs}, the II-CLVM framework achieves an $R@k$ improvement of 0.5$\%$ to 3$\%$ compared to CLVM across various sampling methods. With respect to different encoders, as presented in Table \ref{tab:videomusic}(a), the groups incorporating II loss consistently exhibit superior $R@k$ performance in the video-to-music retrieval task. Although our primary objective is to select music for videos, music-to-video retrieval results are also presented in Table \ref{tab:videomusic}(b). In both retrieval tasks, the proposed framework II-CLVM significantly outperforms previous works \cite{18:18} \cite{19:19}, achieving state-of-the-art $R@k$ results.

\subsubsection{\textbf{Subjective evaluation of video-to-music retrieval}}

A subjective evaluation is necessary for the video music retrieval task. This experiment is conducted similar to \cite{19:19}. Each set of samples consists of one video and two pieces of music. A subject needs to choose a better music from the two candidates for the given video. These materials are all from 642 randomly selected videos from the test data. For each video as query, its $G$, $S$, and $R$ music are choosen in different ways. $G$ (Ground Truth) is the video's original music. $S$ (Select) is the top1 music retrieved by the model biLSTM(II-CLVM) in Table \ref{tab:videomusic}(a). $R$ (Random) is a randomly selected piece of music. For each video, three sets of testing music sample pairs $G-R, G-S, S-R$ are constructed. For set $G-R$, $p_{G>R}$(the probability of $G$ being selected) is calculated. Similarly, $p_{G>S}$ is calculated for set $G-S$ and $p_{S>R}$ for set $S-R$. 

To ensure the credibility of the subjective evaluation results, each set of samples is evaluated by three subjects. Each video is randomly chopped off at the beginning and end to prevent subjects from evaluating by the duration and to make them focus on the global match of video and music. Besides, the file names are all encrypted to prevent subjects from judging by video IDs. 

\begin{table}
\centering
\caption{Results of the subjective evaluation.}
\begin{tabular}{|lll|}
\hline
$p_{G>R}$ & $p_{G>S}$ &  $p_{S>R}$ \\
\hline
88.52$\%$ & 57.27$\%$ &  82.07$\%$ \\
\hline
\end{tabular}
\label{tab:subject}
\end{table}

From Table \ref{tab:subject}, the result of $G-R$ shows that the video and music from the same video ID match very closely. It is feasible to train the matching model by recalling video ID. $p_{G>S}$ close to 50$\%$ indicates that subjects confuse how well $G$ and $S$ match the video. The result of set $S-R$ shows that the BGM selected by the model matches the video better than a random piece of music. In general, our matching model acts great on BGM selection.

\subsubsection{\textbf{II-CLVTM}}

\begin{table}[t]
\caption{Performance comparison of CLTM, CLVM, CLVTM with II-CLTM, II-CLVM, II-CLVTM.}
\centering
\begin{tabular}{|l|lll|}
\hline
Group  & $R@1$ & $R@10$ & $R@25$ \\
\hline
CLTM(keyword) & 5.7$\%$  & 17.9$\%$  & 25.8$\%$  \\
II-CLTM(keyword) & \bf{6.7$\%$}   & \bf{18.2$\%$}  & \bf{26.6$\%$} \\
\hline
CLTM(title) & 6.9$\%$  & 18.8$\%$  & \bf{27.6$\%$}\\
II-CLTM(title) & \bf{7.0$\%$}   & \bf{18.9$\%$}  & 26.9$\%$ \\
\hline
CLVM & 16.6$\%$   & 49.0$\%$  & 65.4$\%$\\
II-CLVM & \bf{20.0$\%$}  & \bf{51.1$\%$}  & \bf{66.1$\%$}  \\
\hline
CLVTM(keyword) & 19.2$\%$   & 52.9$\%$  & 68.6$\%$  \\
II-CLVTM(keyword) & \hl{\bf{22.2$\%$}}   & \hl{\bf{56.6$\%$}} & \hl{\bf{74.0$\%$}} \\
\hline
CLVTM(title) & 18.7$\%$   & 54.5$\%$  & 70.6$\%$  \\
II-CLVTM(title) & \hl{\bf{23.1$\%$}}   & \hl{\bf{60.4$\%$}} & \hl{\bf{75.8$\%$}} \\
\hline
\end{tabular}
\label{tab:iiclvtm}
\end{table}

This subsection evaluates the performance of the II-CLVTM framework. We create a subset of YouTube8M music videos that have both titles and keywords. The subset contains 119,191 video samples, of which 92,604 are for training and 26,587 are for testing. Besides using the official video and music features from YouTube8M, we also extract Clip-text \cite{24:24} feature vectors for titles and keywords. We randomly generate a test set with 1000 sample pairs from the test data. In Table \ref{tab:iiclvtm}, we compare the $R@k(k=1,10,25)$ of II-CLTM(Inter-Intra Contrastive Learning for Text-to-Music Retrieval), II-CLVM, II-CLVTM with CLTM, CLVM, and CLVTM. The results show that using video and text together for music retrieval achieves higher $R@k$ than using video or text alone, regardless of whether the text information is the title or the keyword. Therefore, it is better to select BGM based on multi-modal features rather than a single modality. Moreover, II Loss performs well under all query conditions, especially when multi-modal information is used as the query.

\subsection{II loss on other retrieval tasks} \label{sec:otherexp}

\begin{table}[ht]\footnotesize  
\caption{Cross-modal retrieval datasets used in the experiment. The NOTE column shows the number of sentences for each sample.}
\begin{tabular}{|llll|}
\hline
Dataset  & training data & testing data & NOTE  \\
\hline
\multicolumn{4}{c}{Audio-Text} \\
\hline
CLOTHO\cite{40:40clotho} & \makecell{development (2,981)} & \makecell{evaluation(1,000)} & 5 \\ 
\hline
\multicolumn{4}{c}{Image-Text} \\
\hline
MSCOCO \cite{34:34} & \makecell{train2017(113,287)} & \makecell{val2017(5,000)} & 5\\
Flickr30K \cite{41:41flickr} & \makecell{train(29,783)} &  \makecell{test(1,000)} & 20\\
\hline
\multicolumn{4}{c}{Video-Text} \\
\hline
MSVD \cite{38:38msvd} & \makecell{train(1,200)} & \makecell{test(670)} & 40  \\
MSRVTT \cite{39:39msrvtt} & \makecell{train9k(9,000)} & \makecell{JSFusion(1,000)} & 20\\
VATEX \cite{49:49vatex} & \makecell{train-release(25,991)} & \makecell{test-public(6,000)} & 10 \\
\hline
\end{tabular}
\label{tab:otherdata}
\end{table}

This section evaluates the generality of II loss on different cross-modal retrieval tasks using the audio-text dataset Clotho, the image-text datasets MSCOCO and Flickr30K, and the video-text datasets MSVD, MSRVTT and VATEX. The details of each dataset are shown in Table \ref{tab:otherdata}. In these supervised retrieval tasks, each image, video or audio has multiple corresponding texts(The number of texts is given in the NOTE column of Table \ref{tab:otherdata}). But the training data still contains noise. For instance, in the MSCOCO training set, images with IDs 226419, 166798, 303404 etc. depict planes in the blue sky, and images with IDs 368402, 505728, 509149 etc. portray women in the kitchen. These image-text pairs with similar contents can introduce noise to training. 

The experimental environment and training parameters are the same as \ref{sec:youtubeexp}. BiLSTM is used as the encoder for video and audio feature sequences. For text features, FC layer is used as the encoder. $R@1$ is calculated to test each group of models.  It is worth mentioning that In each epoch, only one text is randomly selected from the multiple texts associated with each image, video or audio sample to form a positive pair for training.

Table \ref{tab:a2t}, \ref{tab:i2t}, \ref{tab:v2t} shows the $R@1$ results of these cross-modal retrieval task. Results for all experimental groups are measured at the steady state of each model after 30 epochs of training. Although there is no longer just one-to-one matches in the datasets of these cross-modal tasks, the noise of false negatives still exists. On CLOTHO dataset, II Loss significantly improves $R@1$ from 2$\%$ to 4$\%$. On the two datasets of image-text retrieval, II Loss improves $R@1$ from 1$\%$ to 2$\%$. On MSVD dataset of video-text retrieval, II Loss improves $R@1$ greater than 5$\%$. And on other datasets of video-text retrieval, II Loss also improves $R@1$ from 0.5$\%$ to 2$\%$. In the meantime, II loss works when different pretrained features are used.

\begin{figure*}[t]
\centering
\includegraphics[scale=0.37]{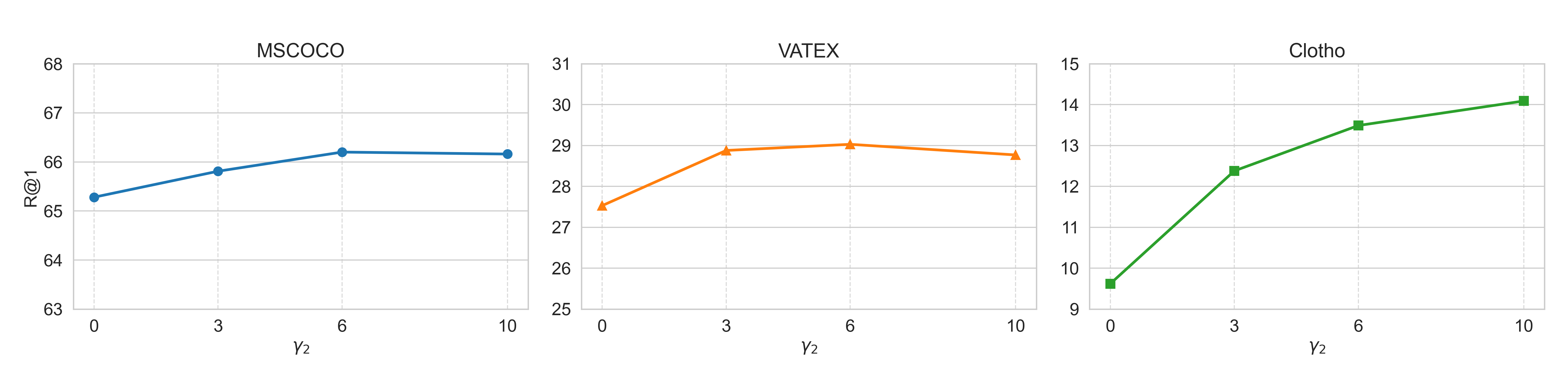} %
\caption{The change curve of $R@1$ as the intra loss weight $\gamma_2$ increases.}
\label{fig:gamma}
\end{figure*}

\begin{table}[ht]\Large
\caption{$R@1$ results of audio-text retrieval.}
\centering
\subfloat[CLOTHO dataset.]{
\resizebox{\columnwidth}{!}{
\begin{tabular}{|l|ll|ll|}
\hline
Features  & inter & inter-intra & inter & inter-intra \\
\cline{2-5}
& \multicolumn{2}{c}{audio-to-text} & \multicolumn{2}{c}{text-to-audio}  \\
\hline
VGGish\cite{29:29}  BERT\cite{35:35} & 5.7$\%$ & \bf{7.7$\%$} & 4.7$\%$ & \bf{6.4$\%$} \\
\hline
PANNs\cite{30:30}  CLIP-text\cite{24:24} & 9.0$\%$ & \bf{13.5$\%$} & 7.9$\%$ & \bf{11.5$\%$} \\
\hline
PANNs\cite{30:30}  BLIP-text\cite{48:48blip} & 9.6$\%$ & \bf{12.4$\%$} & 8.1$\%$ & \bf{10.1$\%$}\\
\hline
\end{tabular}
}}
\label{tab:a2t}
\end{table}

\begin{table}[ht]\Large
\caption{$R@1$ results of image-text retrieval.}
\centering
\subfloat[MSCOCO dataset.]{
\resizebox{\columnwidth}{!}{
\begin{tabular}{|l|ll|ll|}
\hline
Features  & inter & inter-intra & inter & inter-intra \\
\cline{2-5}
& \multicolumn{2}{c}{image-to-text} & \multicolumn{2}{c}{text-to-image}  \\
\hline
ViT-B/32\cite{32:32}  BERT\cite{35:35} & 16.4$\%$ & \bf{17.7$\%$} & 13.8$\%$ & \bf{14.6$\%$}\\
\hline
ViT-B/32\cite{32:32}  CLIP-text\cite{24:24} & 28.1$\%$ & \bf{29.1$\%$} & 21.5$\%$ & \bf{22.1$\%$} \\
\hline
CLIP-ViT\cite{24:24}  CLIP-text\cite{24:24} & 47.0$\%$ & \bf{49.5$\%$} & 35.7$\%$ & \bf{36.0$\%$}\\
\hline
BLIP-image\cite{48:48blip}  BLIP-text\cite{48:48blip} & 65.3$\%$ & \bf{65.8$\%$} & 50.3$\%$ & \bf{51.7$\%$}\\
\hline
\end{tabular}
}}

\subfloat[Flickr30K dataset.]{
\resizebox{\columnwidth}{!}{
\begin{tabular}{|l|ll|ll|}
\hline
Features  & inter & inter-intra & inter & inter-intra \\
\cline{2-5}
& \multicolumn{2}{c}{image-to-text} & \multicolumn{2}{c}{text-to-image}  \\
\hline
ViT-B/32\cite{32:32}  BERT\cite{35:35} & 28.6$\%$ & \bf{30.7$\%$} & 20.8$\%$ & \bf{22.7$\%$}\\
\hline
ViT-B/32\cite{32:32}  CLIP-text\cite{24:24} & 54.6$\%$ & \bf{56.4$\%$} & 43.9$\%$ & \bf{45.9$\%$} \\
\hline
CLIP-ViT\cite{24:24}  CLIP-text\cite{24:24} & 74.7$\%$ & \bf{75.2$\%$} & 61.4$\%$ & \bf{63.0$\%$}\\
\hline
BLIP-image\cite{48:48blip}. BLIP-text\cite{48:48blip} & 82.4$\%$ & \bf{84.7$\%$} & 69.5$\%$ & \bf{71.6$\%$}\\
\hline
\end{tabular}
}}
\label{tab:i2t}
\end{table}

\begin{table}[ht]\Large
\caption{$R@1$ results of video-text retrieval.}
\centering
\subfloat[MSVD dataset.]{
\resizebox{\columnwidth}{!}{
\begin{tabular}{|l|ll|ll|}
\hline
Features  & inter & inter-intra & inter & inter-intra \\
\cline{2-5}
& \multicolumn{2}{c}{video-to-text} & \multicolumn{2}{c}{text-to-video}  \\
\hline
CLIP-ViT\cite{24:24}  BERT\cite{35:35} & 17.8$\%$ & \bf{19.1$\%$} & 11.1$\%$ & \bf{13.3$\%$} \\
\hline
CLIP-ViT\cite{24:24}  CLIP-text\cite{24:24} & 27.1$\%$ & \bf{33.0$\%$} & 23.0$\%$ & \bf{27.1$\%$}\\
\hline
VidSwin\cite{47:47vidswin}  CLIP-text\cite{24:24} & 27.5$\%$ & \bf{32.3$\%$} & 22.6$\%$ & \bf{24.8$\%$}\\
\hline
BLIP-image\cite{48:48blip}  BLIP-text\cite{48:48blip} & 33.2$\%$ & \bf{38.7$\%$} & 26.7$\%$ & \bf{28.1$\%$}\\
\hline
\end{tabular}
}}

\quad
\subfloat[MSRVTT dataset.]{
\resizebox{\columnwidth}{!}{
\begin{tabular}{|l|ll|ll|}
\hline
Features  & inter & inter-intra & inter & inter-intra \\
\cline{2-5}
& \multicolumn{2}{c}{video-to-text} & \multicolumn{2}{c}{text-to-video}  \\
\hline
CLIP-ViT\cite{24:24}  BERT\cite{35:35} & 16.8$\%$ & \bf{17.1$\%$} & 15.5$\%$ & 15.5$\%$ \\
\hline
CLIP-ViT\cite{24:24}   CLIP-text \cite{24:24} & \bf{28.2$\%$} & 27.9$\%$ & 30.5$\%$ & \bf{31.2$\%$} \\
\hline
VidSwin\cite{47:47vidswin}  CLIP-text\cite{24:24} & 22.3$\%$ & \bf{22.6$\%$} & 22.8$\%$ & \bf{23.7$\%$}\\
\hline
BLIP-image\cite{48:48blip}  BLIP-text\cite{48:48blip} & \bf{32.2$\%$} & 31.5$\%$ & 33.1$\%$ &\bf{33.3$\%$} \\
\hline
\end{tabular}
}}

\quad
\subfloat[VATEX dataset.]{
\resizebox{\columnwidth}{!}{
\begin{tabular}{|l|ll|ll|}
\hline
Features  & inter & inter-intra & inter & inter-intra \\
\cline{2-5}
& \multicolumn{2}{c}{video-to-text} & \multicolumn{2}{c}{text-to-video}  \\
\hline
CLIP-ViT\cite{24:24} BERT\cite{35:35} & 13.1$\%$ & \bf{13.1$\%$} & 9.1$\%$ & \bf{9.6$\%$}\\
\hline
CLIP-ViT\cite{24:24}  CLIP-text\cite{24:24} & 28.1$\%$ & \bf{29.5$\%$} & 21.5$\%$ & \bf{22.3$\%$}\\
\hline
VidSwin\cite{47:47vidswin}  CLIP-text\cite{24:24} & 25.1$\%$& \bf{25.9$\%$}& 20.7$\%$ & \bf{21.5$\%$}\\
\hline
BLIP-image\cite{48:48blip}  BLIP-text\cite{48:48blip} & 27.5$\%$ & \bf{28.9$\%$} & 20.6$\%$ & \bf{22.1$\%$}\\
\hline
\end{tabular}
}}
\label{tab:v2t}
\end{table}

\begin{figure*}[ht]
 \centering
 \begin{subfigure}{0.32\linewidth}
 \includegraphics[width=\linewidth]{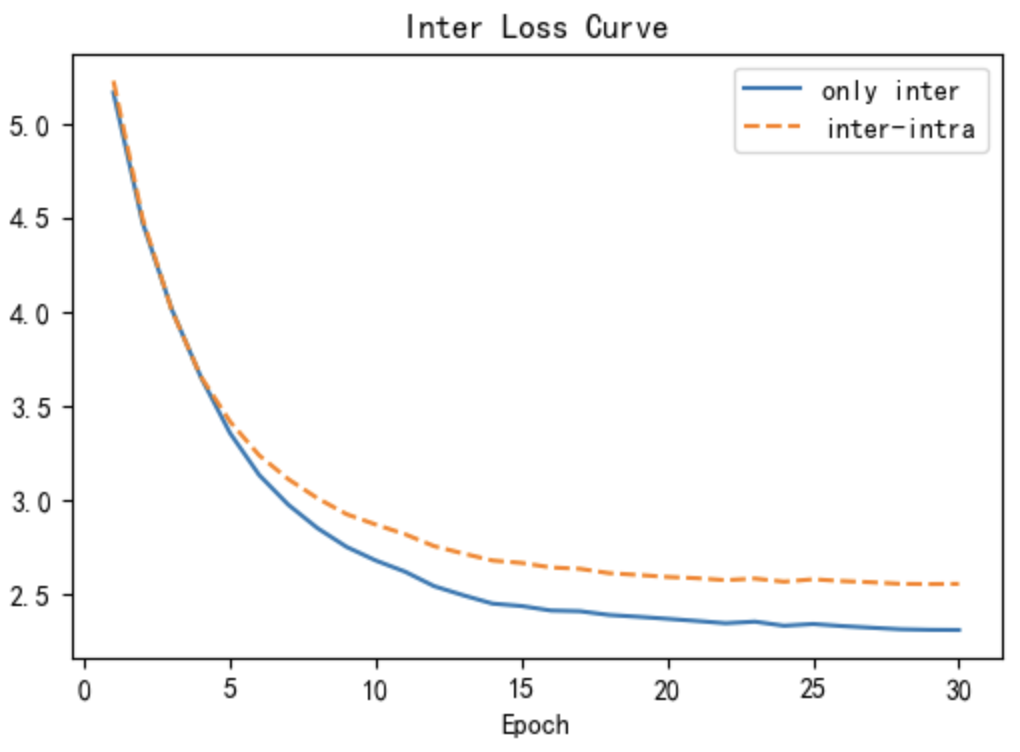}
 \caption{Inter Loss vs Epoch}
 \label{subfig:a}
 \end{subfigure}
 \begin{subfigure}{0.32\linewidth}
 \includegraphics[width=\linewidth]{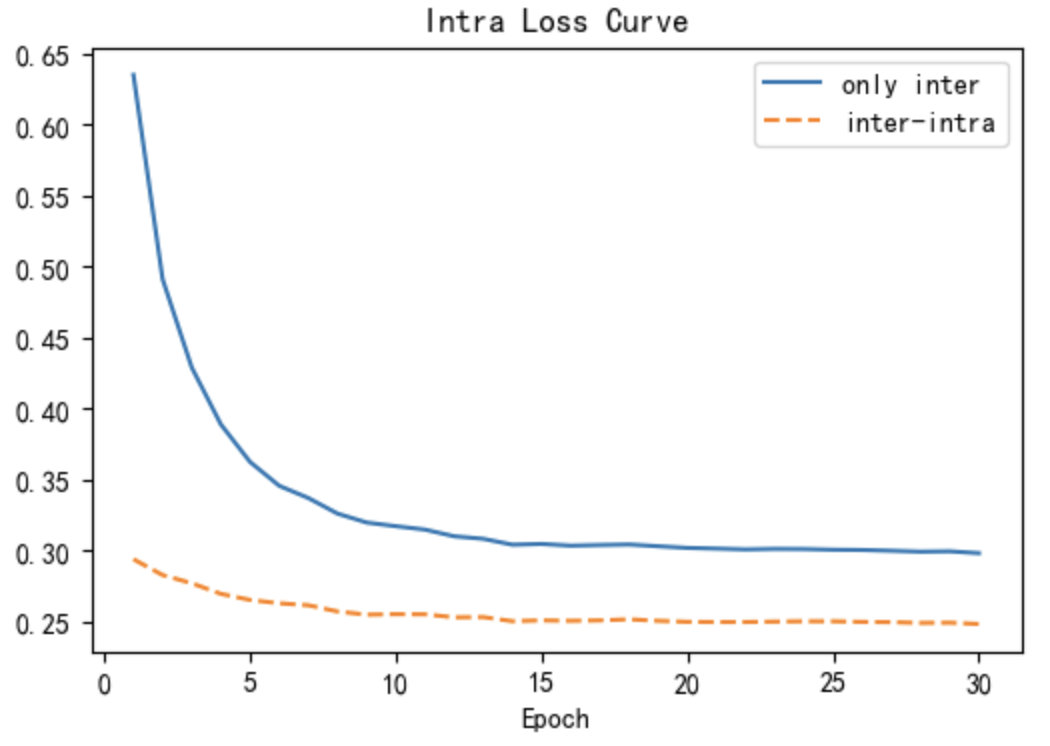}
 \caption{Intra Loss vs Epoch}
 \label{subfig:b}
 \end{subfigure}
 \begin{subfigure}{0.32\linewidth}
 \includegraphics[width=\linewidth]{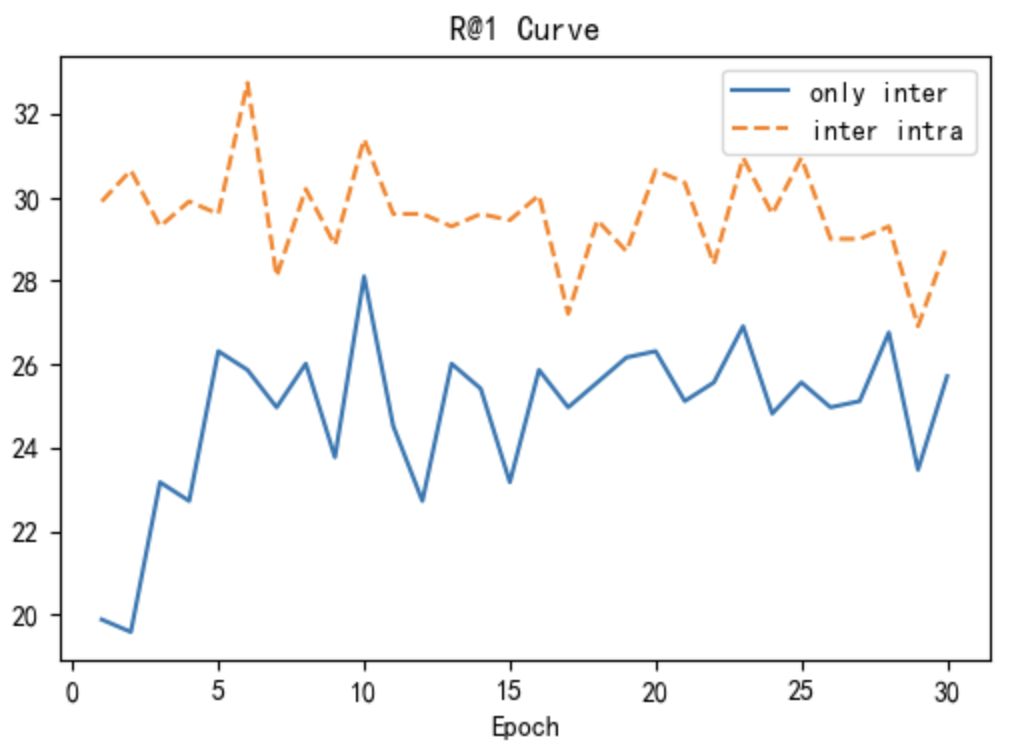}
 \caption{R@1 vs Epoch}
 \label{subfig:c}
 \end{subfigure}
 \caption{The variation curves of the training inter loss, the training intra loss, and the testing $R@1$ over epoch, under the conditions of training only with inter loss and training with the ii loss, respetively.}
 \label{fig:curve}
\end{figure*}

Using the MSVD dataset as an example, the three line charts in Fig \ref{fig:curve} show the inter loss (\ref{subfig:a}) and intra loss (\ref{subfig:b}) during training, and the $R@1$ (\ref{subfig:c}) during testing as a function of the training epoch. The blue solid line represents the group that only uses inter loss, and the orange dashed line represents the group that uses inter-intra (II) loss. Note that in the only-inter group, the intra loss is calculated and plotted in the line chart \ref{subfig:b}, but not back-propagated. As shown in Fig \ref{subfig:a}, compared with the only-inter group, the inter-intra group reached a platform where it could not continue to decrease at a higher inter loss value, suggesting that intra loss can prevent inter loss from overfitting to the noisy training set. In Fig \ref{subfig:b}, Although the intra loss of the inter-intra group is always lower than that of the only-inter group, it does not decrease to a very low value. This is caused by the interaction between inter loss and intra loss. Fig \ref{subfig:c} demonstrates that the $R@1$ score on the testing set for each epoch is higher in the inter-intra group than in the only-inter group.

The three line charts in Fig \ref{fig:gamma} show the trend of $R@1$ indicator on three datasets: MSCOCO, VATEX, and Clotho, as the inter loss weight ($\gamma_1$) takes the value of 1 and the intra loss weight ($\gamma_2$) takes values of 0, 3, 6, and 10. It is obvious that the impact of the II loss is more pronounced in downstream tasks with smaller data volumes. For example, in the Clotho dataset, the $R@1$ values increase from 9.62 to 14.09 as $\gamma_2$ goes from 0 to 10. It is also worth noting that $\gamma_2$ has a noticeable effect on the model's performance. As $\gamma_2$ takes on values of 0, 3, 6, and 10, the $R@1$ indicator generally exhibits an upward trend. The trend slows down gradually and reaches a peak. For smaller datasets (such as clotho), the peak comes later. This is because on small datasets, the model tends to overfit to noisy sample pairs, which gives II loss more room for improvement. This distinctive characteristic of the II loss allows retrieval tasks to achieve satisfactory training results with only a limited number of cross-modal samples collected. To determine the optimal $\gamma_2$ value, a limited number of experiments should be conducted. By analyzing these experiments and the specific $R@1$ values, researchers can select the best intra loss weight to maximize the performance of the model on various datasets and tasks.

\subsection{The effect of II loss against false negative noise interference} \label{sec:thirdexp}

\begin{figure*}[ht]
 \centering
  \begin{subfigure}{0.32\linewidth}
 \includegraphics[width=\linewidth]{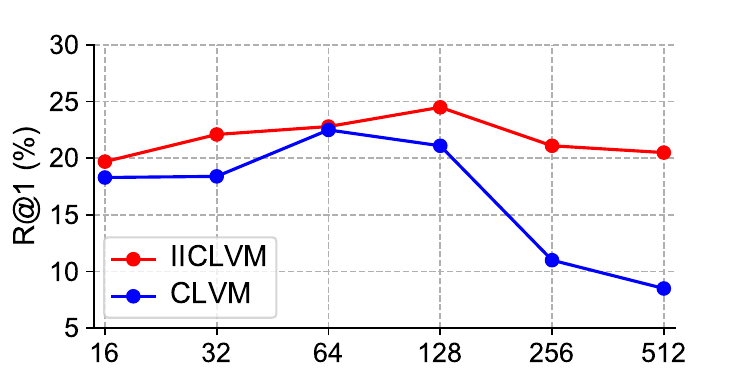}
 \caption{R@1}
 \end{subfigure}
\begin{subfigure}{0.32\linewidth}
 \includegraphics[width=\linewidth]{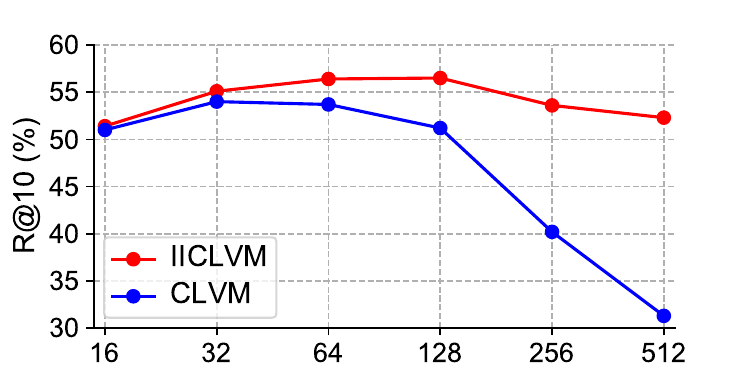}
 \caption{R@10}
 \end{subfigure}
\begin{subfigure}{0.32\linewidth}
 \includegraphics[width=\linewidth]{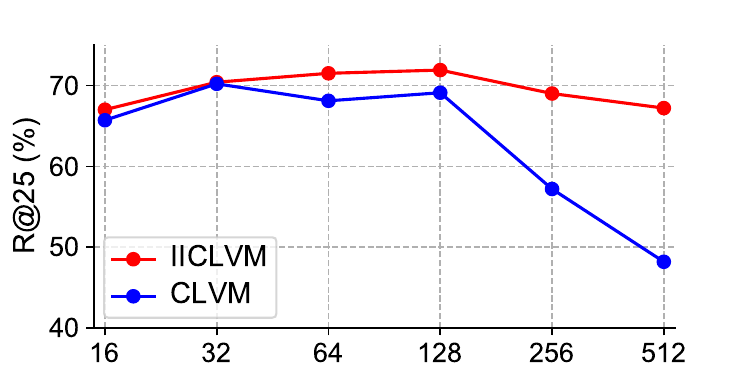}
 \caption{R@25}
 \end{subfigure}

 \caption{The variation curve of $R@k$ changing with batch size on YouTube8M dataset. In CLVM, increasing batch size tends to decrease $R@k$ due to the increased noise. The intra Loss in II-CLVM can reduce the influence of the increased noise and significantly slow down the reduction of $R@k$.}
 \label{fig:batch}
\end{figure*}

This section presents two experiments that focus on examining whether II loss can effectively resist the noise that exists in self-supervised retrieval tasks. The first experiment was performed on the YouTube8M dataset. Based on all the other experimental settings of the group biLSTM(II-CLVM) in section 4.1, this experiment changed the proportion of noise matching pairs by changing the batch size of training, and thus checked the different optimization effects of II loss under different amounts of noise. In Fig \ref{fig:batch}, the $R@k$ metric fluctuates as the batch size changes within the YouTube8m dataset. Initially, the $R@k$ value increases with the growing batch size due to more sufficient contrastive learning. However, as the batch size becomes larger, the number of noisy matching pairs increases exponentially, thereby negatively impacting training and leading to a decline in the $R@k$ value. The effectiveness of II loss in dealing with noise is demonstrated when comparing the II-CLVM group to the CLVM group; the former experiences a considerably slower decrease in $R@k$, suggesting that the II loss is effective in mitigating the adverse effects of noise, and ultimately achieves the highest $R@k$ with a batch size of 128.

\begin{figure*}[ht]
 \centering
  \begin{subfigure}{0.32\linewidth}
 \includegraphics[width=\linewidth]{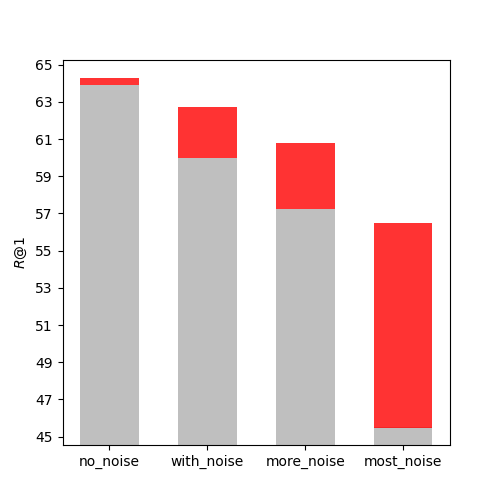}
 \caption{R@1}
 \end{subfigure}
\begin{subfigure}{0.32\linewidth}
 \includegraphics[width=\linewidth]{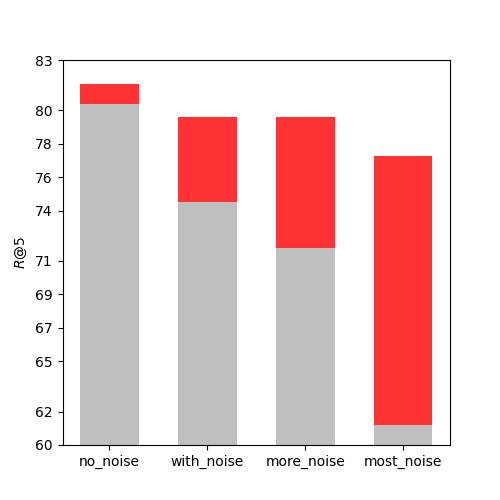}
 \caption{R@5}
 \end{subfigure}
\begin{subfigure}{0.32\linewidth}
 \includegraphics[width=\linewidth]{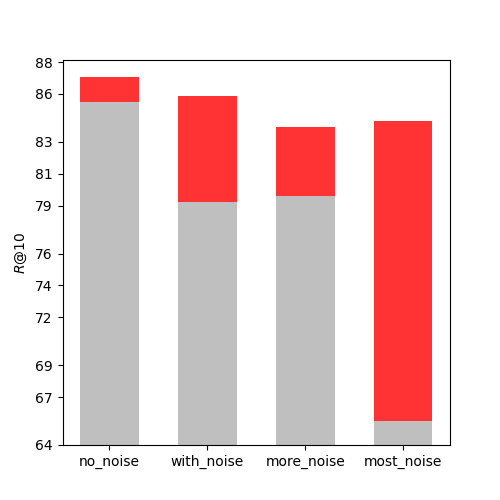}
 \caption{R@10}
 \end{subfigure}

 \caption{A stacked bar chart that shows the relationship between the increment of $R@k$ with or without II loss and the noise level in the batch on dataset Caltech-256 \cite{60:60-256}. The chart uses gray to indicate the $R@k$ without II loss. Red represents the increment of $R@k$ with II loss. As the noise level in the batch increases, the $R@k$ without II loss drops rapidly. The effect of II loss on improving $R@k$ is significant when the noise level is high.}
 \label{fig:imgretrieval}
\end{figure*}

To test the effect of II loss on resisting different levels of noise interference in training, another experiment was conducted on an image classification dataset Caltech-256 \cite{60:60-256} by manually adjusting the noise amount in training batches. The Caltech-256 dataset contains 256 categories of images, with about 100 images per category. The image retrieval task on Caltech-256 was defined as follows: given a query image, the aim was to retrieve another image from the same category as the query image, and $R@k$ was used to measure the retrieval performance. We randomly chose two images from each of the 256 categories as a positive sample pair for the testing set, and used the rest of the images as training data. During training, each pair of positive samples was randomly selected from the same category of training data.  We also arranged multiple pairs of samples from the same category in each training batch to simulate different levels of noise. Four different experiments were performed: no\_noise, with\_noise, more\_noise, and most\_noise, with increasing amounts of noise. N represented the batch size of training. In the no\_noise group, the N pairs of positive samples were selected from N different categories. In the with\_noise group, we picked one pair from each of $\frac{N}{3}$ different categories and two pairs from each of another $\frac{N}{3}$ different categories. In the more\_noise group, we picked two pairs from each of $\frac{N}{2}$ classes. In the most\_noise group, we picked four pairs from each of $\frac{N}{4}$ classes. In this experiment, $N=128$ and $a=6$.  As shown in Fig \ref{fig:imgretrieval}, when there was almost no noise in the no\_noise group, II loss had very little effect on R@k, and even decreased it. As the noise increased, II loss group showed more and more improvement compared to non-II loss group. In real-world self-supervised retrieval datasets, all sample pairs are unlabeled, so there is a high chance that a batch contains similar positive sample pairs. These similar positive sample pairs can produce false negative pairs as mentioned above, which can lower the model’s retrieval performance. II loss is an effective method to enhance the model’s generalization ability under such circumstances.

\section{Conclusion}

The paper emphasizes the II loss as a key innovation in the new framework II-CLVM designed specifically for video-music retrieval. This innovative loss function improves the model's generalization capabilities by maintaining pretrained feature distribution within the two modalities during training on noisy cross-modal datasets. The II-CLVM framework, which incorporates the II loss, has shown promising results in video-music retrieval tasks. The state-of-the-art results achieved on the youtube8m dataset for video-music retrieval tasks demonstrate the effectiveness of the II loss. The II-CLVTM framework with added multi-modal video information input (such as text) has better music retrieval performance in applications. Subjective evaluations also support the framework's strong performance in music selection.

Beyond its application in the II-CLVM framework for video-music retrieval, the II loss has proven to be valuable for other cross-modal retrieval tasks as well, such as image-text, audio-text, and video-text retrieval. This showcases the adaptability and versatility of the II loss in different contexts and applications. Moreover, II loss is found to obtain good retrieval models with a small number of training samples.

Although the current findings are encouraging, the authors acknowledge there is potential for further enhancement, particularly in areas such as retrieval performance on tasks involving large datasets and the development of noise-resistant methods for end-to-end retrieval models. By addressing these challenges, the II loss has the potential to become an even more effective and versatile tool in the field of cross-modal retrieval.

\bibliographystyle{IEEEtran}
\bibliography{tran}

\clearpage
\newpage

\vfill

\end{document}